\documentstyle[twocolumn,aps]{revtex}

\begin{document}
\draft
\newtheorem{theorem}{\indent Theorem}
\newtheorem{corollary}{\indent Corollary}
\newtheorem{lemma}{\indent Lemma}

\title{\hskip10cm {\it NIKHEF--H 94--26, ULB--TH 94/16}
General solution of the Wess-Zumino consistency condition
for Einstein gravity}

\author{Glenn Barnich$^1$, Friedemann Brandt$^2$
and Marc Henneaux$^{1,3}$}

\address{~$^{1}$ Facult\'e des Sciences, Universit\'e Libre de Bruxelles,
Campus Plaine C.P. 231, B-1050 Bruxelles. \\
{}~$^{2}$NIKHEF--H, Postbus 41882,
NL--1009 DB Amsterdam, The Netherlands.  \\
{}~$^{3}$Centro de Estudios Cient\'{\i}ficos de Santiago, Casilla
16443, Santiago 9, Chile.}

\maketitle

\begin{abstract}
The Wess-Zumino consistency condition
for four-dimensional Einstein gravity is investigated
in the space of local forms
involving the fields, the
ghosts, the antifields and their derivatives.  Its
general solution is constructed for
all values of the form degree and of the ghost number.  It is
shown in particular that the antifields (= sources for the BRST
variations) can occur only through cohomologically trivial terms.
\end{abstract}


\narrowtext

The Wess-Zumino (WZ) consistency condition  \cite{wz} is a central equation
in field theories with a local gauge symmetry.  At ghost number
zero, this equation characterizes the consistent deformations of the
original classical action, i.e., its solutions exhaust all possible local
perturbation terms that can be added to the action while maintaining the
number (but not necessarily the form nor the algebra) of the original
gauge invariances \cite{bh}.  An example is given by the Yang-Mills cubic
vertex, which is contained in a ghost number zero solution of
the WZ consistency equation for a system of free abelian gauge
fields.  At ghost number one - for
which it was initially discussed -, the WZ consistency
condition constrains severely the form of the candidate anomalies.
Finally, as it has been shown in \cite{bbh1}, this
condition yields at negative ghost number the characteristic cohomology of the
field equations \cite{Bryant}, which includes the non
trivial conserved currents.

We present in this letter the general solution of the WZ
consistency condition for pure four-dimensional Einstein gravity ,
for arbitrary values of the form degree and of the
ghost number.  Couplings to matter and gravity in other spacetime dimensions
can be handled along the same lines by our method \cite{bbh2}
but are not treated here for the sake of briefness.  We shall just comment
on these generalizations at the end of the letter,
as well as on the modifications
that arise if one adds to the Hilbert-Einstein Lagrangian higher
powers of the curvature.
For simplicity, we assume that the
spacetime manifold ${\cal M}$ has the topology of $R^4$ and
work in global coordinates $x^\mu$ ($\mu = 0,1,2,3$)
throughout.  Our approach does
not use power counting (which would be meaningless in the case
of gravity) and is purely cohomological.

In terms of the BRST operator $s$ given below,
the WZ consistency condition reads
explicitly
\begin{eqnarray}
sa + db = 0  \label{WZ}
\end{eqnarray}
where $a$ and $b$ are local differential forms and where
$d$ is the spacetime exterior derivative.
 Trivial solutions of
(\ref{WZ}) are given by $a = sm
+ dn$ and fulfill indeed (\ref{WZ}) due to $s^2 = sd +ds = d^2 = 0$.
The investigation of (\ref{WZ}) for
gravity has already received considerable attention in the past, particularly
at ghost number one (anomalies)\cite{gravlit} but also at
ghost number zero (deformations of the Hilbert-Einstein action) \cite{wald},
as well as for all ghost numbers \cite{grav}.
Our analysis goes
beyond previous works on the subject in that we do not impose
any a priori form to the dependence on the antifields.  This means,
for instance, that we allow in principle for
deformations of the Hilbert-Einstein action such that
the original off-shell closed gauge algebra is replaced by a new,
possibly off-shell open, gauge algebra.  It turns out,
however - and this is one of our
main results - that all solutions of (\ref{WZ})
involve the antifields only through
trivial terms.  Hence, our results
justify the simplifying assumptions made in previous studies.

The relevant algebra for the discussion of (\ref{WZ})
is the algebra of local differential forms.  This algebra is
the tensor algebra ${\cal E} =  \Omega ({\cal M}) \otimes {\cal A}$, where
$\Omega ({\cal M})$ is the algebra of exterior forms on the manifold ${\cal M}$
and
where ${\cal A}$ is the algebra of local functions.  A local function
is by definition a function of the (invertible)
tetrad $e^{\;\; a}_\mu$ ($a = $ local
index, $\mu = $ world index), of the diffeomorphism ghosts $\xi^\mu$,
of the Lorentz (or $SO(4)$ in the Euclidean
case) ghosts $C^{ab} = - C^{ba}$, of the corresponding
antifields $e^{*\mu}_a$, $\xi^*_\mu$ and $C^*_{ab}$, and of their
derivatives up to some finite order.  Thus, here, local functions
do not depend on $x^\mu$; the explicit $x^\mu$-dependence
is included in the factor $\Omega ({\cal M})$ of
${\cal E}$.  As standard in field theory, we shall assume
that the local functions in ${\cal A}$ depend polynomially on
all their arguments, except on the undifferentiated tetrad,
which may appear non polynomially (e.g. the inverse tetrad $E_a^\mu$
is allowed).

In the algebra of local differential forms, the BRST differential
$s$ acting on $e^{\;\; a}_\mu,\xi^\mu, C^{ab}, e^{*\mu}_a, \xi^*_\mu$ and
$C^*_{ab}$ is obtained by taking the antibracket with the solution $S$
of the master equation \cite{BV,HT}.  In the case of gravity,
whose gauge
symmetries form a closed algebra, the construction of
$S$ presents no difficulty.  One finds that $S$ is given by the sum
of the Einstein action and of integrated
terms of the form ``antifields" times ``BRST variation of
the corresponding fields".  We leave the details to the reader - the
explicit form of $s$ is given below in terms
of a different set of variables.

In order to analyse (\ref{WZ}), it turns
out indeed to be more convenient to
express the local functions in terms of new variables, which are:
\begin{eqnarray}
& &\{{\cal T}^r\}=\{ D_{a_1}\ldots D_{a_m}R_{ab}^{\;\;\; I}:\, m=0,1,\ldots\},
\label{b1}\\
& &\{\hat\xi^a,\hat C^I\},\quad \hat\xi^a=\xi^\mu{e_\mu}^{a},\quad
\hat C^I=C^I+\xi^\mu \omega_\mu^{\; I},
\label{b2}\\
&&  \{{\cal T}^*_r\}=\{D_{a_1}\ldots D_{a_m}\hat \Phi^*_A
:\, m=0,1,\ldots \},\label{b3}\\
& &\{U_l\}=\{ \partial_{(\mu_1 \ldots \mu_m} e_{\mu)}^{\;\; a}, \,
\partial_{(\mu_1 \ldots \mu_m} \omega_{\mu)}^{\;\; I} :\, m=0,1,\ldots \},
\label{b4}\\
& &\{V_l\}=\{sU_l \}.
\label{b5}
\end{eqnarray}
Here (i) the index $I$ collectively denotes the
antisymmetric pair  of Lorentz indices
$ab$, $k^I \equiv k^{ab} = -k^{ba}$ (we use the summation convention
$k^I k_I = k^{ab} k_{ab}/2$); (ii) $\omega^{\; I}_\mu$ is the
standard torsion-free spin connection;
(iii) the antifield variables $\hat\Phi^*_A \equiv
\hat e^*_a{}^b,\hat C^*_I, \hat \xi^*_a$ are defined by
\begin{eqnarray}
&&\hat e^*_a{}^b={e_\mu}^{b}e^{*\mu}_a/e,\label{e*} \\
&&\hat C^*_I=C^*_I/e,\quad
\hat \xi^*_a={E_a}^\mu(\xi^*_\mu- \omega^{\;I}_\mu C^*_I)/e
\label{b6}
\end{eqnarray}
with $e =  \det e^{\;\; a}_\mu$; (iv) $R_{ab}^{\;\;\; I} =
{E_a}^\mu{E_b}^\nu {R_{\mu\nu}}^{I}$ are the tetrad components
of the curvature tensor $R^{\;\;\; I}_{\mu\nu} =
\partial_\mu \omega^{\; I}_\nu - \ldots$;
and  (v)$D_a={E_a}^\mu(\partial_\mu-\omega_{\mu}^{\;\; I} M_I)$. The
symbol $M_I \equiv M_{ab}$ acting on a variable
belonging to a representation of the Lorentz group
stands for its Lorentz variation; it
fulfills
$[M_I, M_J] = f^{\;\;\; K}_{IJ} M_K$,
where $f^{\;\;\; K}_{IJ}$ are the structure constants of the Lorentz
algebra. One easily
verifies that each local function of the fields, the antifields
and their derivatives can indeed be expressed as a local function of the
variables (\ref{b1})--(\ref{b5}) and vice versa.

The BRST transformations of the $U_l$ and $V_l$ are extremely
simple since they just read $sU_l=V_l$, $sV_l=0$.
Accordingly, the $U_l$ and $V_l$ belong to the contractible
part of the algebra and do not contribute to the BRST cohomology.
The BRST transformations of
$\hat \xi^a,\hat C^I$, ${\cal T}^r$ and ${\cal T}^*_r$ are slightly
more complicated.
To display them, we
decompose the BRST operator according to
\begin{equation} s=\delta+\gamma\label{b7}\end{equation}
where $\delta$ is the Koszul-Tate differential which has antighost number $-1$
and $\gamma$ has antighost number 0.  This decomposition is by now
quite standard \cite{FH,Jaca}.  The differential $\delta$
plays a crucial role below and
acts on $\hat \xi^a,\hat C^I$, ${\cal T}^r$, ${\cal T}^*_r$ according to
\begin{eqnarray} & &\delta\, {\cal T}^r=\delta\, \hat C^I=\delta\,
\hat \xi^a=0,\;
\delta\, \hat C^*_{ab}=-2\hat e^*_{[ab]}, \,\nonumber\\
& &\delta\, \hat e^*_a{}^b=
- {R_a}^b+ \frac 12\delta_a^bR, \,\;
\delta\, \hat \xi^*_a=-D_b\hat e^*_a{}^b,\nonumber\\
& &\delta\, D_{a_1}\ldots D_{a_n}\hat \Phi^*_A=D_{a_1}\ldots D_{a_n}\delta\,
\hat \Phi^*_A\label{b8}\end{eqnarray}
where  $R_{ab}=R_{acb}{}^c$ is the
Ricci tensor and $R$ is the scalar curvature.  The differential
$\gamma$ acts on $\hat \xi^a,\hat C^I$, ${\cal T}^r$, ${\cal T}^*_r$
according to
\begin{eqnarray} & &\gamma{\cal T}^r=(\hat \xi^a D_a+\hat C^I M_I){\cal T}^r,\
\gamma{\cal T}^*_r=(\hat \xi^a D_a+\hat C^I M_I){\cal T}^*_r,\nonumber\\
& &\gamma \hat \xi^a= \hat C^I M_I
\hat \xi^a,\quad
\gamma \hat C^I=\frac 12\, \hat C^J\hat C^K{f_{KJ}}^I+\hat R^I\label{b9}
\end{eqnarray}
where we have set $\hat C^I M_I
\hat \xi^a = \hat C_b{}^a\hat \xi^b$ and $\hat R^I=(1/2)
\hat \xi^a\hat \xi^b R_{ab}^{\;\;\; I}$.  The variables ${\cal T}^r$ and
${\cal T}^*_r$, whose BRST variations do not involve derivatives
of the ghosts, have been called ``tensor fields" in \cite{ten}.

  Our first result is :

\begin{theorem}
: The general solution of the Wess-Zumino consistency condition involves
the antifields only through trivial contributions.
\end{theorem}

In order to prove this theorem, we shall need two crucial properties of the
BRST differential.  The first is the fact that the Koszul-Tate
differential provides a resolution of the algebra of on-shell functions.  It
is actually a generic feature of the antifield formalism \cite{FH,Jaca}.  The
second is peculiar to theories with diffeomorphism invariance and relates in a
simple way the cohomology $H(s/d)$ of $s$ modulo $d$
to the cohomology $H(s)$ of $s$ itself
(BRST invariance condition $sa=0$) \cite{grav,ten}.  We shall describe
these two features in turn.  We  shall then provide the demonstration of
the theorem.

Antifields were introduced by Zinn-Justin in order to control
how the non linear BRST symmetry survives renormalization.  They were
known in that context as sources for the BRST variations.  It turns
out that a different interpretation of the antifields is of
greater interest for cohomological investigations.  It is that the
antifields provide a {\em resolution} of the algebra of functions
on the stationary surface \cite{FH,Jaca}, even if
one takes locality into account \cite{CMP}.  This property means :
(i) that an antifield independent local function can be written
as a $\delta$-variation if and only if it vanishes on-shell; and (ii)
that the homology of $\delta$ is trivial at strictly
positive antighost number,
$H_k(\delta) = 0$ for $k>0$.
While the first property is well known and somewhat trivial, the
second property is more subtle and holds because the antifields
associated with the ghosts properly take care of the
relations among the equations of motion (Noether identities).

For pure gravity, the acyclicity of $\delta$ extends to the cohomology of
$\delta$ modulo $d$, which has been shown in \cite{bbh1} to be
isomorphic to the characteristic cohomology
associated with the equations of motion.
Thanks to general theorems valid for generic gauge systems \cite{Bryant,bbh1},
$H_k(\delta/d)$ vanishes for $k>2$ in the case of gravity.  It was
also shown in \cite{bbh1} that $H_2(\delta/d)$ vanishes for gravity because
there is no global reducibility identity among the gauge
transformations.  Finally, $H_1(\delta/d)$, which is isomorphic to the
space of non trivial conserved currents, vanishes in pure gravity
because of a remarkable result due to Anderson and Torre \cite{Torre}.
Thus, one has also $H_k(\delta/d) = 0$ for $k>0$.  We
shall need below a Lorentz invariant refinement of this result, namely,
that if $\delta {\cal L} + \partial_\mu j^\mu = 0$ where
${\cal L} = e L({\cal T}, {\cal T}^*)$ is a Lorentz
invariant density of antighost number $k>0$, then
${\cal L} = \delta m + \partial_\mu k^\mu$, where $m$ has
the same covariance properties.  This refinement is fully proved in
the more complete version \cite{bbh2} of this paper.

The second property needed for the proof of the theorem relates
solutions of $sa+db = 0$ to solutions $\alpha \in {\cal A}$ of
$s\alpha = 0$.  Consider a non trivial cohomological class of $s$ in
${\cal A}$, i.e., a zero-form $\alpha \not= s \beta$  solving
$s \alpha = 0$ (with $\alpha \sim \alpha + s \beta$).  Without loss
of generality, one may assume that $\alpha$ depends only on the tensor
fields ${\cal T}$ and ${\cal T}^*$, and on the undifferentiated ghosts
$\hat \xi, \hat C$.
One can construct from $\alpha$ solutions of $sa + db = 0$ in
two different ways.

\noindent
(a) Because $sx^\mu = sdx^\mu = 0$, any form $\omega \in \Omega({\cal M})$
times $\alpha$ fulfills $s(\omega \alpha) = 0$, and thus, is a
solution of (\ref{WZ}) with $b = 0$.  This solution is trivial if
$\omega = d \eta$.

\noindent
(b) Let $\tilde s = s + d$, $\tilde \xi^a = \hat \xi^a + e^a_\mu dx^\mu$
and $\tilde C^I = \hat C^I + \omega^I_\mu dx^\mu$.  It is easy to
verify that the action of $\tilde s$ on ${\cal T}$, ${\cal T}^*$,
$\tilde \xi$ and $\tilde C$ is exactly the
same as the action of $s$ on ${\cal T}$, ${\cal T}^*$,
$\hat \xi$ and $\hat C$.  Thus, $\tilde s \tilde \alpha = 0$, where
$\tilde \alpha$ is the multi-form obtained by replacing $\tilde \xi$
for $\hat \xi$ and $\tilde C$ for $\hat C$ in $\alpha$.  Each
component of $\tilde \alpha$ of definite form degree $k$ ($= 0,1 ,2,3,4$)
is a solution of (\ref{WZ}) which is not
trivial because $\alpha$ itself is non trivial.

As shown in \cite{grav,ten}, (a) and (b) yield the most
general solution of (\ref{WZ}).  That is,
any solution of (\ref{WZ}) is a linear
combination of solutions of type (a) and of solutions of type (b).
Therefore, in order to solve (\ref{WZ}), it is enough to compute
the cohomology of $s$ in ${\cal A}$. Notice that
all non trivial $4$-forms solving (\ref{WZ}) are of type (b), since
any volume form $\omega \in \Omega ({\cal M})$ is exact.  Therefore the
descent equations \cite{Zumino} in gravity go all the way from
form degree $4$ to form degree $0$ \cite{comment2}.

In order to compute $H(s)$,
we decompose $s$ according to its degree in the Lorentz ghosts
$\hat C^I$,
$s = s_{-1} + s_0 + s_1$,
with
\begin{eqnarray}
&& s_1 \hat C^I = \frac{1}{2} \hat C^J \hat C^K {f_{KJ}}^I,\; \;
s_1 Y = \hat C^I M_I Y
\label{s1}, \\
&& s_0 = \delta + {\cal D} \label{szero}, \\
&& s_{-1} \hat C^I = \hat R^I, \; \;  s_{-1} Y = 0 \label{sminus}.
\end{eqnarray}
Here, we have set ${\cal D} \hat \xi^a = {\cal D} \hat C^I = 0$,
${\cal D\, T}^r = \hat \xi^a D_a {\cal T}^r$, ${\cal D \, T}^*_r =
\hat \xi^a D_a {\cal T}^*_r$ and $Y \equiv ({\cal T,T}^*, \hat \xi)$.
The differential $s_1$ increases the number of Lorentz ghosts by
one unit and is just the standard coboundary operator for the
Lie algebra cohomology of the Lorentz algebra $so(3,1)$, whose cohomology
is well known.  The differential $s_0$ does not modify the number
of Lorentz ghosts, while the differential $s_{-1}$ decreases it
by one unit.

Let $\alpha$ be a solution of $s \alpha = 0$.  We may
also decompose it according to its powers in the Lorentz ghosts,
$\alpha = \alpha_0 + \ldots + \alpha_N$.  From $s \alpha = 0$, one
gets $s_1 \alpha_N = 0$, i.e., $\alpha_N$ is a cocycle
of the Lie algebra cohomology.  Up to $s_1$-trivial terms
that can be absorbed by redefinitions, the
most general $s_1$-cocycle is given by
$\alpha_N = \sum P_\tau ({\cal T, T}^*, \hat \xi) \omega^\tau (\hat C)$,
where the polynomials $P_\tau$ are Lorentz-invariant functions
of their arguments ($M_I P_\tau = 0$) and
where the $\omega^\tau (\hat C)$ belong to the  basis of the Lie algebra
cohomology of $so(3,1)$ explicitly given by $\{\omega^\tau\} =
\{1,\hat \theta_1, \hat\theta_2, \hat\theta_1 \hat\theta_2 \}$,
where $\hat \theta_1
= \eta^{ac} \eta^{bd} \eta^{ef} \hat C_{ab} \hat C_{ce}
\hat C_{df}$ and $\hat \theta_2 = \epsilon^{abcd} \eta^{ef}
\hat C_{ab} \hat C_{ce} \hat C_{df}$.  Thus, $N = 0, 3$ or $6$.

We start with the case $N = 6$.  If one inserts the expression
$\alpha_6 = P ({\cal T, T}^*, \hat \xi) \hat\theta_1 \hat\theta_2$
in the condition $s \alpha = 0$, one gets at $\hat C$-degree
$6$ the equation $(\delta + {\cal D}) P \hat\theta_1 \hat\theta_2 =
s_1$(something), which is possible only if the right and
left hand sides vanish separately ($(\delta + {\cal D}) P$
is a Lorentz invariant polynomial, and no $s_1$-cocycle
$\sum P_\tau \omega^\tau$ of the above form is $s_1$-exact unless
$P_\tau = 0$ for all $\tau$'s).  Thus, $(\delta + {\cal D}) P = 0$.
We analyse this equation by decomposing $P$ according to the
antighost number, $P = P_{0} + \ldots + P_{t}$.  The
condition $(\delta + {\cal D}) P = 0$ implies
${\cal D} P_{t} = 0$.  Assume $t \not= 0$. Since
${\cal D}$ acting on invariant polynomials is just the
exterior derivative operator (after the substitution
$\xi^\mu  \rightarrow dx^\mu$), the covariant
Poincar{\'e} lemma \cite{grav} implies that
$P_{t}$ is ${\cal D}$-trivial in the space of
invariant polynomials in ${\cal T}, {\cal T}^*, \hat \xi$,
unless it contains the
``volume form" $\hat \Theta \equiv \hat \xi^0 \hat \xi^1 \hat \xi^2 \hat \xi^3$
(the substitution $\xi^\mu \rightarrow dx^\mu$ yields $\hat \Theta
\rightarrow d^4x e$).
However, even in that case, the next condition $\delta P_{t}
+ {\cal D} P_{t-1} = 0$ (with $t \not= 0$) implies that $P_{t}$ is
$\delta$-trivial modulo ${\cal D}$ (see above discussion on
the invariant homology $H(\delta/d)$) and can thus be absorbed by
allowed redefinitions.  Accordingly, we
may assume $t = 0$, i.e., $P$ does not contain the
antifields.  The condition ${\cal D} P = 0$ implies then
that $P$ has the form $P = a + \hat \Theta
m({\cal T}^r)$ where $a$ is a
constant (covariant Poincar{\'e} lemma again; ${\cal D}$-exact
pieces can be absorbed by redefinitions).

If $a \not= 0$, $\alpha_6$ cannot be completed to a BRST cocycle.
Indeed, the obstructions for doing this are the characteristic
classes $f_1 = \hat R^{ab} \hat R_{ab}$ and $f_2 = \epsilon^{abcd}
\hat R_{ab} \hat R_{cd}$, which do not vanish even on-shell and can thus not
be written as $\delta$-variations.  Accordingly, we must take
$a = 0$, in which case $\alpha_6 =
\hat \Theta m({\cal T}^r)\hat \theta_1 \hat \theta_2$ is a BRST cocycle.

One repeats the discussion in exactly the same way for the terms of lower
order in the Lorentz ghosts, which must fulfill the
BRST cocycle condition independently of the term of order $6$, since
this one fullfills that condition by itself. For $N=0$, one
gets in addition the constant solutions.  Since
$\hat \Theta$ has maximal degree in the
$\xi^\mu$, one has actually $\hat \Theta \hat \theta_i =
\hat \Theta \theta _i$, where $\theta _i$ is obtained from
$\hat \theta_i$ by replacing $\hat C^I$ by the
ordinary Lorentz ghosts $C^I$.  We can thus conclude:

\begin{theorem}
: Up to $s$-trivial terms, the general solution $\alpha$ of the
BRST cocycle condition $s \alpha = 0$ does not depend on the antifields and
is given by
\begin{equation}
\alpha  = a + \hat \Theta L({\cal T}^r,
\theta_i),\; \hat \Theta = \hat \xi^0 \hat \xi^1 \hat \xi^2 \hat \xi^3
\label{general}
\end{equation}
where $a$ is a constant and $L$ is a Lorentz invariant polynomial.
\end{theorem}
Theorem 1 follows from Theorem 2 since the process by which one
constructs $H(s/d)$ from $H(s)$ does not introduce any
antifield dependence.

One can be even more explicit and list
all the
solutions of (\ref{WZ}).  By the first procedure
(a) described above, one generates from (\ref{general})
solutions
of the form
$ \omega(x^\mu, dx^\mu) \hat \Theta L({\cal T}^r,
\theta_i)$
at ghost numbers $4, 7$ and $10$ and form degrees
$0,1,2,3$.
The
multiform $\tilde \alpha$ obtained by the second procedure (b) is simply
$\tilde \xi^0 \tilde \xi^1 \tilde \xi^2 \tilde \xi^3 L({\cal T}^r, \theta_i)$
and thus the solutions of type (b) read
$(1/k!)dx^{\mu_1}\ldots dx^{\mu_k}\xi^{\mu_{k+1}}\ldots \xi^{\mu_4}
\epsilon_{\mu_1\ldots\mu_4}e L({\cal T}^r, \theta_i)$
at respective form degrees $k = 0,1,2,3,4$ where
$\epsilon_{0123}=1$, $dx^\mu dx^\nu=-dx^\nu dx^\mu\equiv
dx^\mu \wedge dx^\nu$. This
establishes:

\begin{theorem}
: Up to trivial solutions, the
general solution of the Wess-Zumino consistency condition is a linear
combination of solutions of the form
\begin{equation}
\omega(x^\mu, dx^\mu) [ a + \hat \Theta L({\cal T}^r,
\theta_i)]
\end{equation}
and of solutions of the form
\begin{equation}
\frac 1{k!}\, dx^{\mu_1}\ldots dx^{\mu_k}\xi^{\mu_{k+1}}\ldots \xi^{\mu_4}
\epsilon_{\mu_1\ldots\mu_4}\, e \, L({\cal T}^r, \theta_i)
\end{equation}
where $\omega \in \Omega ({\cal M})$ is not
exact, $k = 0,1,2,3,4$ and $L({\cal T}^r, \theta_i)$ is a Lorentz
invariant polynomial in its arguments.  In particular, the
most general local $4$-form solving (\ref{WZ}) is given by
$d^4 x e L({\cal T}^r, \theta_i) + sm + dn$ with
$d^4x = dx^0 dx^1 dx^2 dx^3$.
\end{theorem}

Thus, the only solutions with ghost number zero have form
degree equal to four and are given by Lorentz-invariant polynomials
in the curvature components $R_{abcd}$ and their
successive covariant derivatives
$D_{a_1} \ldots D_{a_m} R_{abcd}$ times $e d^4x$.
These exhaust all the consistent perturbations of the Hilbert-Einstein
action.  Similarly, there is no solution at ghost number $1$ and form
degree 4,
i.e., no gravitational anomaly (in four dimensions), and no
solution at negative ghost number.

In this letter, we have given the explicit list of all the non
trivial solutions of the WZ consistency condition for pure
Einstein gravity in four dimensions.  We have shown in
particular that the dependence on the antifields can
always be removed by adding trivial terms.
Our method applies also if one includes matter couplings or if
one modifies the Hilbert-Einstein Lagrangian by adding higher powers
of the curvature, but there may be then  modifications in
the final results.  Indeed, Yang-Mills fields bring in non
trivial solutions of their own \cite{xxx}.  Furthermore, there may
now exist non trivial
conserved currents (associated e.g. with baryon number conservation).
In that case, $H^{-1}(s/d) \sim H_1(\delta/d)$ does not vanish and
moreover, non trivial antifield-dependent solutions can be
constructed, but only at ghost number $2$ or higher if
the Yang-Mills gauge group is semi-simple.  The discussion
follows closely the pattern developed in \cite{bbh3} and
the details will be reported elsewhere \cite{bbh2}. [We shall
discuss also in that paper the extension to all spacetime dimensions
$\geq 3$, where there are more antifield-independent solutions
- Chern-Simons terms in odd dimensions and
Lorentz anomalies of the Adler-Bardeen type in $2$  mod $4$ dimensions]

M.H. is grateful to E. Witten for his encouragement in pursuing
this problem. G.B. is Aspirant au Fonds National
de la Recherche Scientifique (Belgium). F.B. is supported
by Deutsche Forschungsgemeinschaft.
This work has been partly supported
by a FNRS research grant and by a research contract with the Commission of
the European Communities.


\begin{references}
\bibitem{comment2}  This is in sharp contrast with theories for which the
diffeomorphisms are not gauge symmetries.  For such theories, one
may have (i) nontrivial $4$-form solutions of (\ref{WZ})
that do not descend
to non trivial $0$-forms; and (ii) vice-versa, non-constant
BRST-invariant $0$-forms that cannot be lifted to solutions of (\ref{WZ}) with
higher form degree. When this happens, $d$ is not the zero operator in
$H^*(s)$ and one cannot write
it as a BRST commutator, although this can be done in peculiar
subalgebras \cite{Sorella}. For instance, in
Maxwell theory with charged matter fields, the Lagrangian
(times $d^4x$) provides
an example of (i). Similarly, the individual components
$F_{\alpha \beta}$ of the
field strength exemplify (ii) since they are solutions
of $s F_{\alpha \beta} = 0$, but
 $d F_{\alpha \beta}=
d x^\mu \partial_\mu F_{\alpha \beta} \not= s$(something).
\end{references}
\end{document}